\documentclass[prl,preprint,tightenlines,%
superscriptaddress,showpacs,floatfix,nofootinbib,
]{revtex4}
\usepackage{graphics}

\begin{document}
\preprint{MKPH-T-04-20}
\title{Electromagnetic form factors of the nucleon in chiral
perturbation theory including vector mesons}
\author{Matthias R.~Schindler}
\affiliation{Institut f\"ur Kernphysik, Johannes
Gutenberg-Universit\"at, D-55099 Mainz, Germany}
\author{Jambul Gegelia}
\affiliation{Institut f\"ur Kernphysik, Johannes
Gutenberg-Universit\"at, D-55099 Mainz, Germany} \affiliation{High
Energy Physics Institute, Tbilisi State University, University
St.~9, 380086 Tbilisi, Georgia}
\author{Stefan Scherer}
\affiliation{Institut f\"ur Kernphysik, Johannes
Gutenberg-Universit\"at, D-55099 Mainz, Germany}

\begin{abstract}
We calculate the electromagnetic form factors of the nucleon up to
fourth order in manifestly Lorentz-invariant chiral perturbation
theory with vector mesons as explicit degrees of freedom.
   A systematic power counting for the renormalized diagrams is implemented
using both the extended on-mass-shell renormalization scheme and the reformulated
version of infrared regularization.
   We analyze the electric and magnetic Sachs form factors, $G_E$ and $G_M$, and
compare our results with the existing data. The inclusion of vector mesons
results in a considerably improved description of the form factors.
    We observe that the most dominant contributions come from
tree-level diagrams, while loop corrections with internal vector
meson lines are small.
\end{abstract}

\date{November 9, 2004}
\maketitle


\section{\label{introduction} Introduction}

   Experiments on elastic electron-nucleon scattering provide the most
fundamental information on the electromagnetic structure of the nucleon
\cite{Mcallister:1956ng}.
   In the one-photon-exchange approximation, this information is contained in
four electromagnetic form factors, two each for the proton and the
neutron, which parameterize the single-nucleon matrix element of
the electromagnetic current operator $J^\mu(x)$:
\begin{displaymath}
\label{FFDef} \langle N(p')| J^{\mu}(0) |N(p)\rangle =
   \bar{u}(p')\left[\gamma^{\mu}F_1^N
   +\frac{i\sigma^{\mu\nu}q_{\nu}}{2m_p}F_2^N\right]u(p),
\end{displaymath}
where $q=p'-p$  and $N=p,n$. The so-called Dirac and Pauli form factors
$F_1^N(Q^2)$ and $F_2^N(Q^2)$ are functions of $Q^2=-q^2\ge 0$ and are normalized
such that, at $Q^2=0$, they reduce to the electric charge and the anomalous
magnetic moment in units of the elementary charge and the nuclear magneton
$e/(2m_p)$, respectively:
\begin{displaymath}
F_1^p(0)=1, \,\, F_1^n(0)=0,\,\, F_2^p(0)=1.793, \,\, F_2^n(0)=-1.913.
\end{displaymath}
   For the analysis of experimental data it is more convenient to use the electric and
magnetic Sachs form factors $G_E^N(Q^2)$ and $G_M^N(Q^2)$ \cite{ESW 60} which are
related to the Dirac and Pauli form factors via
\begin{eqnarray*}
G_E^N(Q^2)&=&F_1^N(Q^2)-\frac{Q^2}{4m_N^2}F_2^N(Q^2),\\
G_M^N(Q^2)&=&F_1^N(Q^2)+F_2^N(Q^2).
\end{eqnarray*}
Their Fourier transforms in the Breit frame can be related to the distribution of
charge and magnetization inside the nucleon.
   These form factors have been the aim
of extensive research and, for the case of the proton, are known over a wide
momentum range.
   An apparent inconsistency of the results for the ratio of the electric and magnetic proton
form factors as obtained from the Rosenbluth separation in comparison with those
from the polarization transfer method has recently been addressed in terms of
two-photon exchange corrections \cite{Guichon:2003qm}.
   Due to the lack of a free neutron target, the neutron
form factors are not as well known. However, recent experiments using polarized
beams and/or targets have improved our knowledge, especially of $G_E^n$ (for an
overview and a recent discussion of the existing form factor data, see
Refs.~\cite{Kelly:2002if,Gao:2003ag,Friedrich:2003iz} and references therein).
Given the wealth and precision of available data, the description of the
electromagnetic form factors presents a stringent test for any theory or model of
the strong interaction.

   Chiral perturbation theory (ChPT) \cite{Weinberg:1978kz,Gasser:1984yg,Gasser:1988rb}
is the effective field theory of quantum chromodynamics in the low-energy region
(for a recent review, see Ref.\ \cite{Scherer:2002tk}).
   Using ChPT, the form factors have been calculated
within the early relativistic approach \cite{Gasser:1988rb}, heavy-baryon ChPT
\cite{Bernard:1992qa,Fearing:1997dp}, and the small-scale expansion
\cite{Bernard:1998gv}.
   The spectral functions of the isovector electromagnetic form factors of the nucleon
have been analyzed at the one- and two-loop order in Refs.\ \cite{Bernard:1996cc}
and \cite{Kaiser:2003qp}, respectively.
   More recently, also two manifestly Lorentz-invariant renormalization schemes,
namely infrared regularization (IR) of Ref.~\cite{Becher:1999he} and the extended
on-mass-shell (EOMS) scheme  of Ref.~\cite{Fuchs:2003qc}, have been used to
calculate the form factors up to and including order ${\cal O}(q^4)$
\cite{Kubis:2000zd,Fuchs:2003ir}.
   The results in the two renormalization schemes
are very similar, but fail to describe both proton form factors $G_E^p$ and
$G_M^p$ as well as the magnetic neutron form factor $G_M^n$ for momentum
transfers beyond $Q^2\sim 0.1\, \mbox{GeV}^2$. To improve these results
higher-order contributions have to be included. This can be
achieved by performing a full 
calculation at ${\cal O}(q^5)$ which would also include the analysis of two-loop
diagrams.
   That such a calculation in a manifestly Lorentz-invariant framework is, at
least in principle, possible has been demonstrated in Ref.\
\cite{Schindler:2003je}.

   Another possibility is to include additional degrees of freedom, through which
some of the higher-order contributions are re-summed.
   This latter approach is less systematic and proceeds more along the lines of
phenomenological models.
   It has long been established that vector mesons play an important role in the
description of the nucleon form factors, and by including them dynamically in the
effective field theory one hopes to generate the most important higher-order
contributions. In Ref.\ \cite{Kubis:2000zd} the $\rho$, $\omega$, and $\phi$
mesons have been included in the calculation. One finds that the vector mesons
re-sum important higher-order contributions and the obtained description of form
factors up to $Q^2\approx 0.4\, \mbox{GeV}^2$ is satisfactory. Since diagrams
with internal vector meson lines {\em inside} loops cannot be treated within the
original formulation of infrared regularization, such diagrams have not been
considered in Ref.\ \cite{Kubis:2000zd}.

    The EOMS renormalization scheme of Ref.~\cite{Fuchs:2003qc} and
the reformulated version of infrared regularization of
Ref.~\cite{Schindler:2003xv} both allow to include virtual vector mesons {\em
systematically} in the region of the applicability of baryon chiral perturbation
theory \cite{Fuchs:2003sh}. The standard power counting determines which diagrams
(including diagrams with vector mesons appearing in loops) should be taken into
account to a given order in the chiral expansion.
    In this Letter we present the nucleon electromagnetic form
factors to order ${\cal O}\left( q^4\right)$ in the framework of baryon ChPT with
explicit vector mesons. Our calculations contain {\it all} diagrams which appear
to this order in the chiral expansion.

\section{\label{effective_lagrangian} Effective Lagrangian and Power Counting}
   The Lagrangian needed for calculating the electromagnetic form
factors up to order $q^4$ without explicit vector mesons can be found in
Ref.~\cite{Fuchs:2003ir}.
   Here, $q$ collectively stands for a small quantity such as
the pion mass, small external four-momenta of the pion and small external
three-momenta of the nucleon.
   In this Letter, we consider, in addition the $\rho$, $\omega$, and $\phi$
mesons as explicit degrees of freedom. We make use of the vector-field
representation of Ref.\ \cite{Ecker:yg}, in which the $\rho$ meson is represented
by $\rho_{\mu}=\rho_{\mu}^a\tau^a$ and the $\omega$ and $\phi$ mesons by
$\omega_\mu$ and $\phi_\mu$, respectively. The coupling of the vector mesons to
pions and external fields is at least of the order $q^3$,
\begin{equation}\label{PionLagrange}
   \mathcal{L}_{{\pi}V}^{(3)}=-f_\rho\mbox{Tr}(\rho^{\mu\nu}f^+_{\mu\nu})
-f_\omega\omega^{\mu\nu}f^{(s)}_{\mu\nu}-f_\phi\phi^{\mu\nu}f^{(s)}_{\mu\nu}+\ldots,
\end{equation}
 where
\begin{displaymath}
f^{(s)}_{\mu\nu}=\partial_\mu v^{(s)}_\nu-\partial_\nu
v^{(s)}_\mu,
\end{displaymath}
\begin{displaymath}
f^+_{\mu\nu}=u f^L_{\mu\nu} u^++u^+f^R_{\mu\nu}u,
\end{displaymath}
with
\begin{displaymath}
f^L_{\mu\nu}=\partial_\mu l_{\nu} -\partial_\nu l_\mu -i
\left[l_\mu,l_\nu\right],
\end{displaymath}
\begin{displaymath}
f^R_{\mu\nu}=\partial_\mu r_{\nu} -\partial_\nu r_\mu -i
\left[r_\mu,r_\nu\right].
\end{displaymath}
   The SU(2) matrix $u^2=U$ 
   contains the pion fields.
   For the case of a coupling to an external electromagnetic potential ${\cal A}_\mu$
the external fields are given by $r_\mu=l_\mu=-e\tau_3{\cal A}_\mu/2$ and
$v_\mu^{(s)}=-e {\cal A}_\mu/2$ ($e^2/4\pi\approx 1/137$, $e>0$)
\cite{Scherer:2002tk}.
   Furthermore
\begin{displaymath}
    V_{\mu\nu}=\nabla_{\mu}V_{\nu}-\nabla_{\nu}V_{\mu},\quad
    V=\rho,\omega,\phi
\end{displaymath}
 with
\begin{displaymath}\label{Vkovariant1}
  \nabla_{\mu}V_{\nu}=\partial_{\mu}V_{\nu}+\left[\Gamma_{\mu},V_{\nu}\right]
\end{displaymath}
 and
\begin{displaymath}\label{Gamma}
  \Gamma_{\mu}=\frac{1}{2}\left[u^\dagger,\partial_{\mu}u\right]
-\frac{i}{2}u^\dagger r_\mu u-\frac{i}{2}u l_\mu u^\dagger.
\end{displaymath}
   Only those terms that are used for the calculation of the form
factors up to order $q^4$ are given here; a complete list of possible interaction
terms at order $q^3$ can be found in Ref.\ \cite{Ecker:yg}.

   The lowest-order Lagrangian for the coupling to the nucleon is
given by
\begin{equation}\label{LagNV0}
   \mathcal{L}_{NV}^{(0)}=\frac{1}{2}\sum_{V=\rho,\omega,\phi}g_V\,\bar{\Psi}\gamma^{\mu}V_{\mu}\Psi,
\end{equation}
and the ${\cal O}(q)$ Lagrangian reads
\begin{equation}\label{LagNV1}
   \mathcal{L}_{NV}^{(1)}=\frac{1}{4}\sum_{V=\rho,\omega,\phi}G_V
   \bar{\Psi}\sigma^{\mu\nu}V_{\mu\nu}\Psi.
\end{equation}

  Finally, each renormalized diagram has a chiral order $D$ which is
determined with the following power counting rules in addition to the ones in
Ref.~\cite{Fuchs:2003ir}: vertices from $\mathcal{L}_{{\pi}V}^{(3)}$ count as
$\mathcal{O}(q^3)$ and vertices from $\mathcal{L}_{NV}^{(i)}$ as
$\mathcal{O}(q^i)$, respectively, while the vector-meson propagators count as
$\mathcal{O}(q^0)$.

\section{\label{results} Results and Discussion}

\begin{figure}
\includegraphics{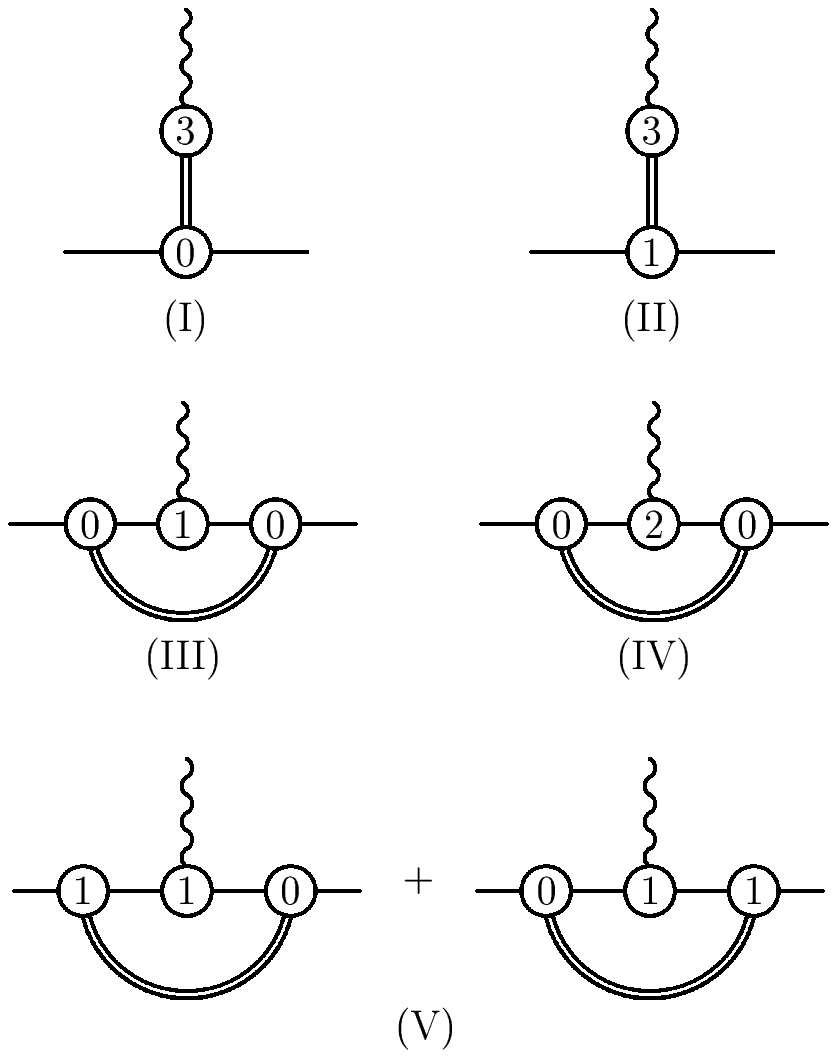}
\caption{\label{Dia_mit} Feynman diagrams
including vector mesons that contribute to the electromagnetic form factors of
the nucleon up to and including ${\cal O}(q^4)$. External leg corrections are not
shown. Solid, wiggly, and double lines refer to nucleons, photons, and vector
mesons, respectively. The numbers in the interaction blobs denote the order of
the Lagrangian from which they are obtained. The direct coupling of the photon to
the nucleon is obtained from ${\cal L}_{\pi N}^{(1)}$ and ${\cal L}_{\pi
N}^{(2)}$.}
\end{figure}

The relevant diagrams that do {\em not} contain vector mesons and their explicit
contributions to the form factors are given in Ref.~\cite{Fuchs:2003ir}. The
additional diagrams involving vector mesons that contribute in the calculation of
the form factors up to and including order ${\cal O}(q^4)$ using the Lagrangians
of Eqs.\ (\ref{PionLagrange}), (\ref{LagNV0}), and (\ref{LagNV1}) are shown in
Fig.~\ref{Dia_mit}. To renormalize the results we employed both the infrared
regularization \cite{Becher:1999he} in its reformulated version
\cite{Schindler:2003xv} as well as the EOMS scheme of Ref.~\cite{Fuchs:2003qc}.
   Since the results for the vector-meson diagrams contain only loop integrals with
no internal pion lines, these loop contributions vanish in the infrared
regularization and only the tree graphs (I) and (II) of Fig.~\ref{Dia_mit}
contribute.
   On the other hand, in the EOMS scheme the renormalized diagram (III) contributes
at ${\cal O}(q^3)$ and the renormalized diagrams (IV) and (V) at ${\cal O}(q^4)$,
respectively.
   Finally, at the given order the vector-meson diagrams do not contribute to the
wave function renormalization constant $Z$.

   In order to obtain the form factors numerically
we have to fix the parameters of the Lagrangian.
   The parameters of the vector-meson Lagrangian of Eq.\ (\ref{PionLagrange})
for the coupling to external fields have been taken from Ref.~\cite{Ecker:yg},
and those of Eqs.\ (\ref{LagNV0}) and (\ref{LagNV1}) for the coupling of vector
mesons to the nucleon from the dispersion relations of
Refs.~\cite{Mergell:1995bf,Hammer:2003ai}.
   The numerical values of these coupling constants are given
   in Table~\ref{FFKonstfg}.
\begin{table}
\caption{\label{FFKonstfg} Values of the vector-meson coupling
   constants within an ${\cal O}(q^4)$ calculation.}
\begin{ruledtabular}
\begin{tabular}{ccccccccc}
$f_\rho$&$f_\omega$&$f_\phi$&$g_\rho$&$g_\omega$&$g_\phi$&$G_\rho$&
$G_\omega$&$G_\phi$\\
&&&&&& [GeV$^{-1}$]&[GeV$^{-1}$]&[GeV$^{-1}$]\\
   0.10&0.03&0.05&4.0&42.8&$-20.6$&13.0&0.96&$-3.3$
\end{tabular}
\end{ruledtabular}
\end{table}

   To determine the low-energy constants (LECs) $c_2$, $c_4$,
$\tilde{c_6}$, $\tilde{c_7}$, $d_6$, $d_7$, $e_{54}$, and $e_{74}$
of the $\pi N$ effective Lagrangian \cite{Fettes:2000gb} we
proceed analogously to Ref.~\cite{Fuchs:2003ir}. We use the more
recent values of Ref.~\cite{Hammer:2003ai} for the proton electric
and magnetic radii, $r_E^p=0.848\,{\rm fm}$ and $r_M^p=0.857\,{\rm
fm}$, and the neutron magnetic radius, $r_M^n=0.879\,{\rm fm}$.
Table~\ref{FFKonstLEC} summarizes the values of the LECs in the
EOMS and infrared regularization schemes. The differences between
the LECs in the respective renormalization schemes originate in
the different treatment of loop integrals.
   In comparison to a calculation without vector mesons only the parameters $d_i$
and $e_i$ change.

\begin{table}
\caption{\label{FFKonstLEC} Values of the relevant low-energy
   constants in the EOMS and the infrared regularization scheme. The
   LECs $c_i$ are given in units of GeV$^{-1}$, the $d_i$ in units of
   GeV$^{-2}$, and the $e_i$ in units of GeV$^{-3}$.}
\begin{ruledtabular}
   \begin{tabular}{ccccccccc}
   &$c_2$&$c_4$&$\tilde{c_6}$&$\tilde{c_7}$&$d_6$&$d_7$&$e_{54}$&$e_{74}$\\
   EOMS&2.66&2.45&1.26&$-0.13$&1.21&1.30&$-0.76$&1.65\\
   IR&2.66&2.45&0.47&$-1.87$&0.98&0.24&$-0.26$&$-0.90$\\
   \end{tabular}
\end{ruledtabular}
\end{table}

\begin{figure}
\includegraphics{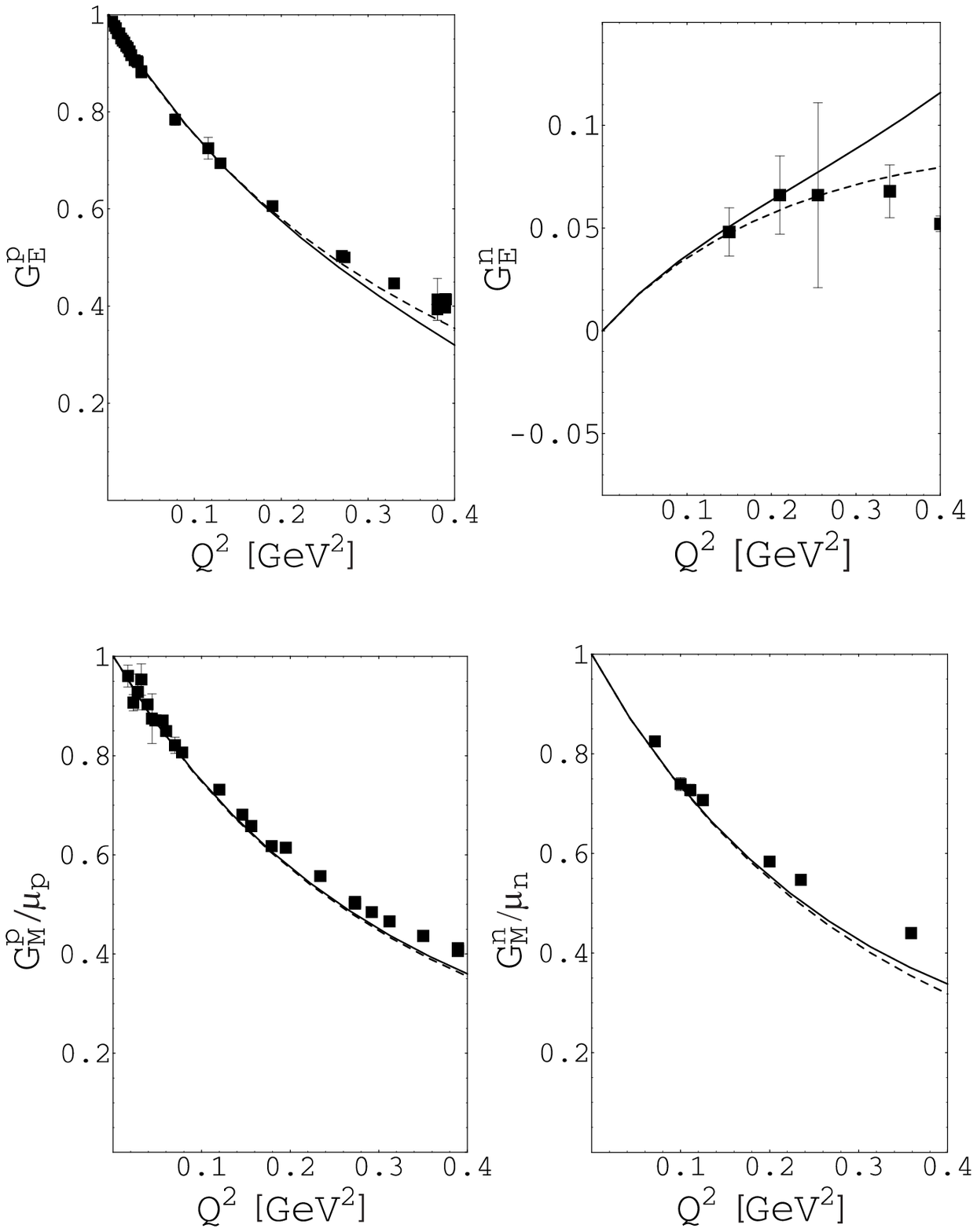}
\caption{\label{G_neu} The Sachs form factors of the nucleon in
manifestly Lorentz-invariant chiral perturbation theory at ${\cal
O}(q^4)$ including vector mesons as explicit degrees of freedom.
Full lines: results in the extended on-mass-shell scheme; dashed
lines: results in infrared regularization. The experimental data
are taken from Ref.~\cite{Friedrich:2003iz}.}
\end{figure}

\begin{figure}
\includegraphics{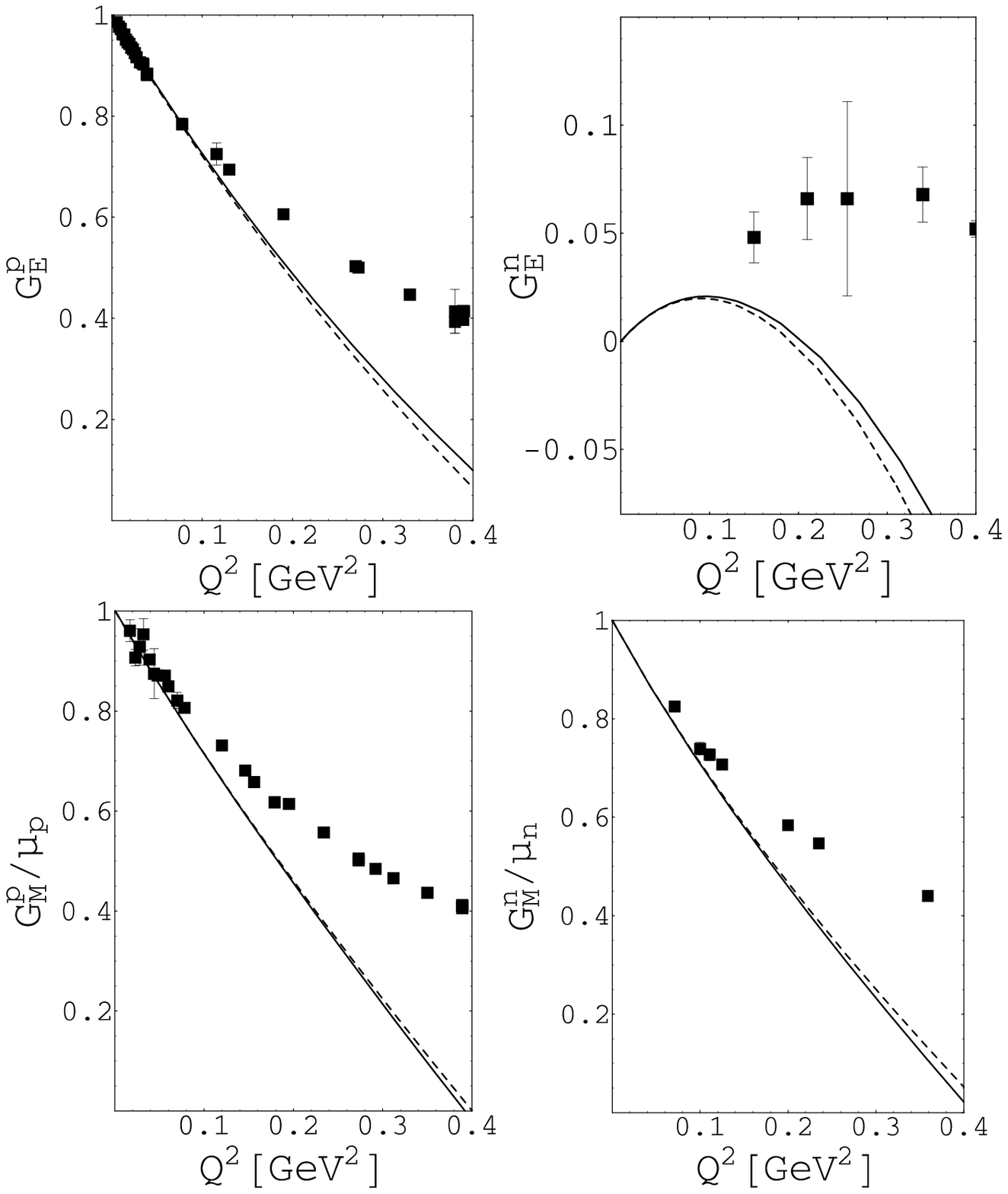}
\caption{\label{G_ohne} The Sachs form factors of the nucleon in
manifestly Lorentz-invariant chiral perturbation theory at ${\cal
O}(q^4)$ without vector mesons. Full lines: results in the
extended on-mass-shell scheme; dashed lines: results in infrared
regularization. The experimental data are taken from
Ref.~\cite{Friedrich:2003iz}.}
\end{figure}

   The results for the Sachs form factors in the momentum transfer region
$0\,{\rm GeV^2}\le Q^2\le 0.4\,{\rm GeV^2}$ are shown in Fig.~\ref{G_neu}.
   For comparison, Fig.~\ref{G_ohne} contains the corresponding results at ${\cal O}(q^4)$
without vector mesons.
   As expected on phenomenological grounds, the quantitative description of the
data has improved considerably for $Q^2\gtrsim 0.1$ GeV$^2$.
   The small difference between the two renormalization schemes is due to the
way how the regular higher-order terms of loop integrals are treated.
   Note that on an absolute scale the differences between the two schemes are
comparable for both $G_E^p$ and $G_E^n$.
   Numerically, the results are similar to those of Ref.\ \cite{Kubis:2000zd}.
   Due to the renormalization condition, the contribution of the vector-meson
loop diagrams either vanishes (IR) or turns out to be small (EOMS).
   Thus, in hindsight our approach puts the traditional phenomenological
vector-meson dominance model on a more solid theoretical basis.
   We would like to emphasize that, in the sense of a strict chiral expansion
in terms of small external momenta $q$ and quark masses $m_q$ at a fixed ratio
$m_q/q^2$ \cite{Gasser:1984yg}, up to and including ${\cal O}(q^4)$ the results
with and without explicit vector mesons are completely equivalent.
   The additional vector-meson contributions up to this order are
compensated by a readjustment of the low-energy constants pertaining to the
theory including vector mesons as dynamical degrees of freedom.
   On the other hand, the inclusion of vector-meson degrees of freedom in the present
framework results in a reordering of terms which, in an ordinary chiral
expansion, would show up at higher orders $q^5$ and $q^6$ etc.
   It is these terms which change the form factor results favorably for larger
values of $Q^2$.
   It should be noted, however, that this re-organization proceeds according to
well-defined rules so that a controlled, order-by-order, calculation of
corrections is made possible.
   In contrast to the calculation without vector mesons, the Sachs form factors
$G_E^p$, $G_M^p$, and $G_M^n$ now show sufficient curvature to
generate a more accurate  phenomenology for values of $Q^2$, where
the ordinary chiral expansion to the same order is no longer
reliable.
   Also the description of $G_E^n$ has improved considerably when
compared to a calculation without the inclusion of vector mesons.

   To conclude, we have shown how the traditional, but ad hoc and
phenomenological vector-meson dominance model can be incorporated consistently
into an effective field theory approach.
   Using a suitable renormalization condition we were able to set up a
systematic power counting and to justify that the vector-meson loop contributions
are suppressed while the tree-level pole diagrams generate the well-known
important contributions to the nucleon electromagnetic form factors.

\acknowledgements We would like to thank J.~Friedrich and Th.~Walcher for
providing us with their data base, and D.~Drechsel and H.~W.~Fearing for useful
comments on the manuscript. J.G.~and M.R.S.~acknowledge the support of the
Deutsche Forschungsgemeinschaft (SFB 443).


\end{document}